\documentclass[epsf,useAMS,usenatbib, usegraphicx]{mn2e}
\usepackage{graphicx}
\usepackage{epsfig}
\usepackage{amsmath}
\usepackage{amssymb}
\usepackage{rotating}
\usepackage{color}
\usepackage{pifont}
\usepackage{longtable}
\usepackage{pdflscape,lscape}
\usepackage{float}
\usepackage{placeins}
\newcommand{\ion}[2]{{#1}\,{\sc #2}}
\newcommand{\teff}{$T_{\rm eff}$}
\newcommand{\logg}{$\log g$}
\newcommand{\vsini}{$v \sin i$}
\newcommand{\kms}{km\,s$^{-1}$}
\usepackage{appendix}
\usepackage{booktabs}
\usepackage{upgreek}
\usepackage{url}

\usepackage{natbib}
\title[High-resolution spectroscopy of $\delta$\,Scuti stars]{High-resolution spectroscopy and abundance analysis of $\delta$\,Scuti stars near the $\gamma$\,Doradus instability strip}

\author[F. Kahraman Ali\c{c}avu\c{s} et al.]{F. Kahraman Ali\c{c}avu\c{s}$^{1,2}$\thanks{E-mail: filizkahraman01@gmail.com/filizkahraman@comu.edu.tr},
E. Niemczura$^{2}$, M. Poli\'{n}ska$^{3}$, K. G. He\l{}miniak$^{4,5}$,\and P. Lampens$^{6}$, J. Molenda-\.{Z}akowicz$^{2,7}$, N. Ukita$^{8,9}$, E. Kambe$^{8}$
\\
$^{1}$Canakkale Onsekiz Mart University, Faculty of Sciences and Arts, Physics Department, 17100, Canakkale, Turkey\\
$^{2}$Instytut Astronomiczny, Uniwersytet Wroc\l{}awski, ul. Kopernika 11, 51-622 Wroc\l{}aw, Poland\\ 
$^{3}$Astronomical Observatory Institute, Faculty of Physics, A. Mickiewicz University, S\l{}oneczna, 36, 60-286, Pozna\'{n}, Poland\\
$^{4}$Subaru Telescope,  National Astronomical Observatory of Japan, Hilo, HI 96720, USA\\
$^{5}$Department of Astrophysics, Nicolaus Copernicus Astronomical Center, ul. Rabia\'nska 8, PL-87-100 Toru\'n, Poland\\
$^{6}$Royal Observatory of Belgium, Ringlaan 3, B-1180 Brussel, Belgium \\
$^{7}$Department of Astronomy, New Mexico State University, Las Cruces, NM 88003, USA\\
$^{8}$Okayama Astrophysical Observatory, National Astronomical Observatory of Japan, 3037-5 Honjo, Kamogata, Asakuchi,\\
Okayama 719-0232, Japan\\
$^{9}$The Graduate University for Advanced Studies, 2-21-1 Osawa, Mitaka, Tokyo 181-8588, Japan\\
}
\begin{document}

\date{Accepted ... Received ...; in original form ...}

\pagerange{\pageref{firstpage}--\pageref{lastpage}} \pubyear{2017}

\maketitle

\label{firstpage}

\begin{abstract}

$\delta$\,Scuti stars are remarkable objects for asteroseismology. 
In spite of decades of investigations, there are still important questions about these pulsating stars to be answered,
such as their positions in $\log$\,\teff\,$-$\,\logg\ diagram, or the dependence of the pulsation modes on atmospheric parameters and rotation. 
Therefore, we performed a detailed spectroscopic study of $41$ $\delta$\,Scuti stars. 
The selected objects are located near the $\gamma$\,Doradus instability strip to make a reliable comparison between both types of variables.
Spectral classification, stellar atmospheric parameters (\teff, \logg, $\xi$) and \vsini\ values were determined.
The spectral types and luminosity classes of stars were found to be A1\,$-$\,F5 and III\,$-$\,V, respectively. 
The \teff\ ranges from $6600$ to $9400$\,K, whereas the obtained \logg\ values are from $3.4$ to $4.3$.
The \vsini\ values were found between $10$ and $222$\,km\,s$^{-1}$.
The derived chemical abundances of $\delta$\,Scuti stars were compared to those of the non-pulsating stars and $\gamma$\,Doradus variables.
It turned out that both $\delta$\,Scuti and $\gamma$\,Doradus variables have similar abundance patterns, which 
are slightly different from the non-pulsating stars. 
These chemical differences can help us to understand why there are non-pulsating stars in classical instability strip. 
Effects of the obtained parameters on pulsation period and amplitude were examined. 
It appears that the pulsation period decreases with increasing \teff.
No significant correlations were found between pulsation period, amplitude and \vsini.

\end{abstract}


\section{Introduction}

Asteroseismology provides a great opportunity to probe the internal structures of stars by modelling their pulsation modes. 
Many pulsating stars have been examined in detail to determine their pulsational properties. 
One of the remarkable objects for asteroseismology are $\delta$\,Scuti ($\delta$\,Sct) variables 
because of their large number of pulsation modes, amplitude regimes, which ranges from low ($\leq 0^{m}.1$) to high ($\geq 0^{m}.3$) amplitudes, 
and their positions in the Hertzsprung-Russell (H-R) diagram. 

The $\delta$\,Sct stars range from dwarf to giant stars and have spectral types between A0 and F5 \citep{2013AJ....145..132C}. 
These stars oscillate in radial and non-radial, low-order pressure (p), gravity (g) and mixed modes excited by the $\kappa$-mechanism.
Most $\delta$\,Sct stars pulsate in the frequency range from $5$ to $50$~d$^{-1}$.
Their masses vary between $1.5$ and $2.5$\,$M_{\rm \odot}$ \citep{2013AJ....145..132C}.
The coolest $\delta$\,Sct stars in the $\delta$\,Sct instability strip are in the transition region,
where the convective envelope gradually turns into a radiative one,
and the energy is transferred by convection in the core \citep[Section 3.7.3]{2010aste.book.....A}. 
Investigations of $\delta$\,Sct stars will help us to understand processes occurring in the transition region.

$\delta$\,Sct variables are located in the lower part of the classical instability strip where 
the theoretical instability strips of $\delta$\,Sct and $\gamma$\,Doradus ($\gamma$\,Dor) variables partially overlap.
In this overlapping area, $\delta$\,Sct/$\gamma$\,Dor hybrids were predicted 
observationally \citep{1996A&A...313..851B, 2002MNRAS.333..251H} and theoretically \citep{2004A&A...414L..17D}. 
The number of these variables increased with the discoveries based on space telescopes observations. 
Thanks to the high precision photometry obtained by the {\sl Kepler} space telescope, many $\delta$\,Sct, 
$\gamma$\,Dor, and candidate hybrid stars have been discovered \citep{2011A&A...534A.125U, 2010ApJ...713L.192G}. 
\citet{2011A&A...534A.125U} showed the position of $\delta$\,Sct, $\gamma$\,Dor, and candidate hybrid stars in the H-R diagram mainly by using the photometric 
atmospheric parameters taken from the {\sl Kepler} input catalog (KIC) \citep{2011AJ....142..112B}. 
It turned out that $\delta$\,Sct and $\gamma$\,Dor stars can be found outside their theoretical instability strips. 
The candidate hybrid stars have been discovered in the instability strip of the $\delta$\,Sct as well as $\gamma$\,Dor stars.
To find out the exact positions of $\delta$\,Sct and the other A-F type pulsating stars, spectroscopic studies are essential.

Some detailed spectroscopic studies dealing with these variables have been carried out
\citep{2011MNRAS.411.1167C, 2013MNRAS.431.3685T, 2015MNRAS.450.2764N, 2016MNRAS.458.2307K}. 
The atmospheric parameters and the abundance patterns of $\delta$\,Sct, $\gamma$\,Dor, 
and hybrid stars were derived and their accurate positions in the $\log$\,\teff\,$-$\,\logg\ have been determined. 
According to these studies, $\delta$\,Sct stars were found mainly inside their theoretical instability strip. 
A detailed abundance pattern of the $\delta$\,Sct stars was investigated by \citet{2008A&A...485..257F}
to check whether the assumption of solar abundance in the pulsation models was correct. 
They found that generally the elements \ion{Y}\,and \ion{Ba}\ are overabundant.
The abundance pattern of $\delta$\,Sct stars were also compared with the non-pulsating A-F type stars and no significant differences were found.
However, it should be kept in mind that only few $\delta$\,Sct stars were used in this study.

On the other hand, a lot of studies based on the frequency analysis of $\delta$\,Sct stars were presented.
For instance, \citet{2011MNRAS.417..591B} and \citet{2014MNRAS.437.1476B} examined the pulsation frequencies of $\delta$\,Sct stars using the {\sl Kepler} data. 
They showed that the $\delta$\,Sct stars beyond the blue edge of their instability strip
pulsate in high-radial and non-radial overtones as suggested by \citet{1975ApJ...200..343B}. 
The authors tested the working hypothesis that rotation in connection with stellar spots (i.e. rotational modulation) could explain the 
low frequencies but conclude that ``rotational splitting, by itself, cannot account for the number of low frequencies nor their distribution''.
In the study of \citet{2011MNRAS.417..591B} a gap between zero age main sequence (ZAMS) 
and the position of $\delta$\,Sct stars in the H-R diagram was also found. 
It was shown that the gap increases with growing effective temperature. 

Although much is already known about their stellar properties, several big issues concerning $\delta$\,Sct stars are still unanswered.
For example, the exact position of $\delta$\,Sct stars in the H-R diagram and borders of their instability strip need to be checked observationally. 
Additionally, $\delta$\,Sct stars in the red border of their instability strip have almost the same atmospheric parameters as $\gamma$\,Dor stars 
but pulsate in different modes.
It is known that chemical composition influences the opacity in stars and opacity is related to the $\kappa$ mechanism.
Additionally, it is known that the chemical composition affects the pulsation modes \citep{2008MNRAS.386.1487M}.
Therefore, this situation could be explained by a possible chemical abundance differences between $\delta$\,Sct and $\gamma$\,Dor stars. 
The other question concerns the relation between the pulsation quantities and rotation and metallicity of $\delta$\,Sct stars.

To answer all these questions detailed spectroscopic studies of $\delta$\,Sct stars are necessary, preferably using high-resolution observations. 
Hence, in this study, we aim to obtain fundamental atmospheric parameters and chemical composition of a sample of $\delta$\,Sct stars
based on high-resolution spectra taken with different instruments. 
We have selected a sample of $41$ $\delta$\,Sct stars from the catalogue of \citet{2000A&AS..144..469R}.
31 of these stars are confirmed $\delta$\,Sct variables and the others are suspected $\delta$\,Sct variables which 
show pulsation but do not have realiable photometric studies to confirm $\delta$\,Sct type variability.

Stars were selected considering their position in the H-R diagram. 
The $\delta$\,Sct stars located in/near the overlapping region of $\delta$\,Sct and $\gamma$\,Dor instability strips were chosen.
In Sect.\,2 details of the observations and data reductions are given.
Spectral classification of stars is presented in the Sect.\,3. 
Determinations of initial atmospheric parameters from photometric indices of several systems and from spectral energy distribution are introduced in Sect.\,4.
In Sect.\,5, the spectroscopic determination of atmospheric parameters and chemical abundances are given. 
Discussion of the results and conclusions are presented in Sect.\,6 and Set.\,7, respectively. 

\section[]{Observations}

In our analysis, we used high-resolution spectra taken with four spectrographs: ARCES, ELODIE, HERMES and HIDES.
The information about the spectroscopic survey is given in Table\,\ref{table1}. 
The signal-to-noise (S/N) ratios of the spectra near the wavelength $5500$\,{\AA} are listed in Table\,\ref{table2}. 

The ARC \'{E}chelle Spectrograph (ARCES) is a high-resolution, cross-dispersed optical spectrograph mounted
at the $3.5$-m telescope of the Apache Point Observatory (USA).
It captures the entire spectrum between $3200$ and $10000$\,\AA\ in a single exposure.
ARCES provides spectra with a resolving power of $\rm{R}\sim31\,500$.
For the reduction of the data, we used the NOAO/IRAF\footnote{http://iraf.noao.edu/} package
and the procedure described in the ARCES data reduction cookbook\footnote{The ARCES Data Reduction Cookbook by Karen Kinemuchi is available at the website 
http://astronomy.nmsu.edu:8000/apo-wiki/wiki/ARCES no1.}. 
The reduction process included the bias subtraction, bad pixels fixing, trimming, scattered light correction,
removal of the cosmic rays, flat-field correction and calibration in wavelength done on the basis of the exposures of the ThAr calibration lamps.
The spectra were extracted with the use of the \textit{apall} task also provided by IRAF.

ELODIE is a cross-dispersed \'{e}chelle spectrograph used at the $1.93$-m telescope of Observatoire de Haute Provence (OHP, France) between late 1993 and mid 2006.
The spectra cover the wavelength range from $3850$ to $6800$\,\AA\ with a resolving power of $42\,000$.
The standard data reduction of the ELODIE data was performed automatically with the dedicated pipeline.
The reduced archival ELODIE data were taken from the public archive\footnote{http://atlas.obs-hp.fr/elodie/} \citep{2004PASP..116..693M}.

The High Efficiency and Resolution Mercator \'{E}chelle Spectrograph (HERMES) is a high-resolution fibre-fed
\'{e}chelle spectrograph attached to the $1.2$-m Mercator telescope at the Roque de los Muchachos Observatory  (ORM, La Palma, Spain) \citep{2011A&A...526A..69R}.
The spectra acquired in the high-resolution fiber have a resolving power $\rm{R}\sim85\,000$ and cover the spectral range from $3770$ to $9000$\,{\AA}.
The data have been reduced with a dedicated pipeline\footnote{http://hermes-as.oma.be/doxygen/html/index.html}, 
which includes bias subtraction, extraction of scattered light, cosmic ray filtering, wavelength calibration by a ThArNe lamps and order merging.

The HIgh-Dispersion \'{E}chelle Spectrograph \citep[HIDES][]{izu99} is attached to the $1.88$-m telescope of the Okayama Astrophysical Observatory (Japan).
The spectra cover the visual wavelength range with the resolving power $\rm{R}\sim50\,000$ with its high-efficiency fiber-link \citep{2013PASJ...65...15K}.
The reduction was made using dedicated IRAF-based scripts that deal with all chips simultaneously, and included bias, flat-field, and scattered light subtraction,
corrections for bad pixels and cosmic rays, aperture extraction, and wavelength calibration.
The last step was done on the basis of ThAr lamp exposures.
Spectra from three chips were later merged into one file.
Due to crowding of the apertures and sudden drop in the signal from the bluest chip, we have extracted only the $24$ reddest orders.
The final product is composed of $62$ spectral orders, spanning from $4080$ to $7520$\,\AA.   
A more detailed description of the data reduction process is given in \citet{2016MNRAS.461.2896H}.

\begin{table}
\centering
  \caption{Information about the spectroscopic survey. N gives the number of observed stars.}
  \label{table1}
  \begin{tabular}{@{}lcccccc@{}}
  \toprule
Instrument  & N     & Observations       & Resolving   & Spectral       \\
            &       &  years             & power       & range [{\AA}]  \\
 \midrule
 ARCES    & 4       & 2015            & 31500   & 3850\,$-$\,10500 \\
 ELODIE   & 8       & 1999\,$-$\,2004 & 42000   & 3900\,$-$\,6800  \\
 HERMES   & 6       & 2015            & 85000   & 3770\,$-$\,9000  \\
 HIDES    & 23      & 2015            & 50000   & 4080\,$-$\,7525  \\
\bottomrule
\end{tabular}
\end{table}

\begin{table*}
\centering
  \caption{Information about the investigated stars.}
  \label{table2}
  \begin{tabular*}{0.9\linewidth}{@{\extracolsep{\fill}}llllcll}
  \toprule
     HD   & Name     & Instrument   &  S/N   & V      & Sp. type    & Sp. type      \\
   number &          &              &        &(mag)   & (Simbad)    & (this study)  \\ 
\midrule
   089843$^{a,1}$ & EN UMa        & ARCES   &  210   &5.90    & A7\,Vn      & A6\,V \\
   099002         & CX UMa        & ARCES   &  190   &6.94    & F0          & A7\,V \\
   115308         & DK Vir        & ARCES   &  220   &6.69    & F1\,IV      & F2\,V \\
                  & $^{*}$EH Lib  & ARCES   &  150   &9.83    & A5          & A7\,V \\
\midrule    
   010088 &                       & ELODIE  &  115   &7.86    & A0          & F0\,V   \\
   012389 &                       & ELODIE  &  100   &7.98    & A0          & A5\,IV  \\
   023156 & V624 Tau              & ELODIE  &  190   &8.22    & A7\,V       & A6\,V   \\
   062437 & AZ CMi                & ELODIE  &  220   &6.47    & F0\,III     & A7\,IV  \\
   073857 & $^{*}$VZ Cnc          & ELODIE  &  70    &7.18    & A9\,III     & F2\,III \\
   103313 & IQ Vir                & ELODIE  &  230   &6.31    & F0\,V       & A6\,V   \\
   110377 & GG Vir                & ELODIE  &  210   &6.22    & A6\,Vp      & A5\,IV  \\
   191635 & $^{**}$V2109 Cyg      & ELODIE  &  140   &7.49    & F0          & F5\,IV  \\
\midrule 
   192871 & $^{*}$V383 Vul        & HERMES  &  180   &7.51    & F1\,V(n)    & F2\,IV \\
   199908 & DQ Cep                & HERMES  &  175   &7.26    & F1\,IV      & F2\,V  \\
   210957 &                       & HERMES  &  160   &8.00    & A9\,IV      & A6\,V  \\
   213272$^{a,2}$ & $^{*}$        & HERMES  &  180   &6.55    & A2\,V       & A1\,IV \\
   214698$^{a,3}$ & 41 Peg        & HERMES  &  200   &6.33    & A2\,V       & A1\,IV \\
   219586$^{a,4}$ & V388 Cep      & HERMES  &  160   &5.55    & F0\,IV      & F1\,IV \\
\midrule   
   034409 & BS Cam           & HIDES        &  100   &8.43    & F2          & F2\,IV \\
   037819 & V356 Aur         & HIDES        &  130   &8.08    & F8          & hF3mF6\,III\\
   050018 & OX Aur           & HIDES        &  90    &6.10    & F2\,Ve      & F2\,IV \\
   060302$^{a,5}$ & V344 Gem & HIDES        &  80    &8.02    & F0          & F0\,IV \\
   079781 & GG UMa           & HIDES        &  100   &8.59    & F5          & hF0mF3\,III \\
   081882 & KZ UMa           & HIDES        &  130   &8.12    & F0          & A9\,V \\
   082620 & DL UMa           & HIDES        &  170   &7.18    & F0          & F2\,V \\
   084800 & IX UMa           & HIDES        &  170   &7.90    & A2\,II      & A1\,V \\
   090747$^{a,6}$ & GS UMa   & HIDES        &  85    &8.66    & F8          & F4\,IV\\
   093044 & EO UMa           & HIDES        &  135   &7.11    & A7\,III     & F0\,V \\                             
   097302 & FI UMa           & HIDES        &  70    &6.64    & A4\,V       & A4\,IV  \\
   099983 & HQ UMa           & HIDES        &  120   &7.10    & F0          & F2\,IV  \\
   102355 & KW UMa           & HIDES        &  80    &6.60    & F0          & A2\,IV  \\
   118954$^{a,5}$ & IP UMa   & HIDES        &  110   &7.66    & A5          & F2\,V   \\
   127411 & IT Dra           & HIDES        &  140   &7.52    & A2          & A2\,V/IV\\                            
   151938 & V919 Her         & HIDES        &  120   &8.35    & F2          & F3\,IV  \\
   154225 & V929 Her         & HIDES        &  120   &7.96    & A5m         & F2\,V   \\
   155118 & V873 Her         & HIDES 	    &  100   &8.40    & F0          & F3\,III \\
   161287 & V966 Her         & HIDES        &  100   &7.98    & F2          & F1\,IV  \\
   176445 & V1438 Aql        & HIDES        &  80    &7.72    & F0          & F3\,III \\
   176503$^{a,5}$ & V544 Lyr & HIDES        &  120   &7.43    & F0          & A7\,IV \\
   184522$^{a,5}$ & V2084 Cyg& HIDES        &  100   &7.34    & A3          & F2\,IV \\
   453111$^{a,5}$ & V456 Aur & HIDES	    &  80    &7.85    & F0          & F2\,III\\
\bottomrule
\end{tabular*}
  \begin{description}
  \item[ ]*High\,-\,amplitude $\delta$\,Sct stars (HADS); **star of intermediate amplitude variation (between HADS and low\,-\,amplitude $\delta$\,Sct stars 
  (LADS) \citep{2013AJ....145..132C}. $^{a}$Suspected $\delta$\,Sct stars. $^{1}$\citet{1988IBVS.3225....1P}, $^{2}$\citet{1991AJ....101.2177S}, $^{3}$\citet{1993PASP..105...22S}, 
  $^{4}$\citet{1995A&A...297..754H}, $^{5}$\citet{1999IBVS.4659....1K}, $^{6}$\citet{1997IBVS.4513....1D}
  \end{description}
\end{table*} 
 
The normalisation of spectra was performed manually using the \textit{continuum} task of the NOAO/IRAF package.
The spectra were divided into several parts and each part was normalised individually.
The normalisation was checked by comparing the observed spectrum with a synthetic spectrum assuming an approximate effective temperature (\teff),
surface gravity (\logg) and projected rotational velocity (\vsini) values.
Because of limited observation time, typically one spectrum was obtained per star.
However, line profiles in each spectrum were checked carefully to be sure whether the spectroscopic double-lined binary (SB2) stars are present in our sample.
A few suspect SB2 stars were found (GW\,Dra, EE\,Cam, V1162\,Ori, and GX\,Peg) and excluded from the further analysis. 
In the case of multiple spectra available for a star, we investigated the averaged spectrum. 
The averaging process was applied after normalisation. 

\section{Spectral classification}

Stellar spectral types and luminosity classes were obtained by using the classical spectral classification method \citep{gray&corbally2009} 
of comparison spectra of the studied stars with those of standards \citep{2003AJ....126.2048G}. 
Our sample consists of A and F stars, so the spectral types are typically derived from hydrogen H$\delta$ and H$\gamma$, \ion{Ca}{ii}\,K, and metal lines.
For non-chemically peculiar (non-CP) objects, all these lines provide the same spectral type. 
These lines are available only in the spectra taken with HERMES.  
In addition, the hydrogen lines are an excellent tool to derive luminosity types of A stars,
but the sensitivity to the luminosity decreases in late A type stars and the ionised metal lines become useful \citep{gray&corbally2009}.
Therefore, the luminosity class was obtained from the hydrogen lines for early A stars, whereas for later types 
the lines of ionised \ion{Fe}{} and \ion{Ti}{} were used.

Spectral types of the analysed stars were derived in the range from A1 to F5, and luminosity classes from III to V. 
For some objects, spectral classification was difficult because of the available spectral range and normalisation problems occurring typically in broad Balmer lines.
The normalisation problem was caused by the merging of short orders in the ARCES and HIDES spectra.
Additionally, the spectral range of HIDES data starts around $4080$\,{\AA},
therefore the analysis of Ca\,II H and K lines, as well as the H$\delta$ line was impossible for the spectra taken with this instrument.
The literature and new spectral classes of the investigated stars are shown in Table~\ref{table2}. 

\begin{table*}
\centering
\footnotesize
  \caption{The interstellar reddening ${\it E(B-V)}$ and derived atmospheric parameters from photometric indices and SED analysis.}
  \begin{tabular*}{0.95\linewidth}{@{\extracolsep{\fill}}clcccccccccc@{}}
\toprule         
     HD    & Name & \textit{E(B-V)}   & \textit{E(B-V)} & \textit{E(B-V)}      &  \teff\        &\logg\               &  \teff\            & \logg\             &\teff\              & \teff\         &\teff\ \\
   number  &      & $^{uvby\beta}$           &  $^{Map}$       & $^{NaD_2}$           & $^{uvby\beta}$ &  $^{uvby\beta}$     &    $^{Geneva}$     &  $^{Geneva}$       & $^{UBV}$           &   $^{2MASS}$   &$^{SED}$          \\
           &      & (mag)             &  (mag)           &    (mag)             & (K)            &                     &    (K)             &                    & (K)                & (K)            &(K)          \\  
           &      &                   &                 &                      &  $\pm$\ 95     & $\pm$\ 0.10         & $\pm$\ 125         & $\pm$\ 0.11        & $\pm$\ 170         & $\pm$\ 80      &\\
\midrule
   089843 & EN UMa  & 0.00 & 0.01 & 0.00 & 7260 & 3.29 &      &      & 7560 & 7260 & 7245\,$\pm$\,120 \\
   099002 & CX UMa  & 0.00 & 0.01 & 0.00 & 7500 & 3.65 & 7480 & 4.08 & 7410 & 7400 & 7550\,$\pm$\,150  \\
   115308 & DK Vir  & 0.00 & 0.00 & 0.00 & 6960 & 3.44 & 6960 & 3.86 & 7100 & 6955 & 7110\,$\pm$\,100 \\
          & EH Lib  &      & 0.02 & 0.00 &      &      &      &      & 7440 & 7250 & 7235\,$\pm$\,100\\
   010088 &         & 0.00 & 0.03 & 0.00 & 7310 & 4.02 & 7480 & 4.38 & 7050 & 7340 & 7000\,$\pm$\,150\\
   012389 &         & 0.08 & 0.04 & 0.05 & 8230 & 3.82 &      &      & 8050 & 7820 & 8300\,$\pm$\,250\\
   023156 & V624 Tau& 0.04 & 0.03 & 0.01 & 7970 & 4.25 & 7700 & 4.47 & 7400 & 7400 & 7720\,$\pm$\,180 \\
   062437 & AZ CMi  & 0.01 & 0.01 & 0.00 & 7820 & 3.61 & 7700 & 3.81 & 7785 & 7575 & 7420\,$\pm$\,150\\
   073857 & VZ Cnc  & 0.00 & 0.02 & 0.01 & 6450 & 2.86 & 7000 & 3.78 & 7210 & 7170 & 7190\,$\pm$\,210\\
   103313 & IQ Vir  & 0.02 & 0.01 & 0.01 & 7870 & 3.68 & 7695 & 3.84 & 7785 & 7585 & 7460\,$\pm$\,130\\
   110377 & GG Vir  & 0.00 & 0.00 & 0.00 & 7720 & 3.89 & 7790 & 4.26 & 7850 & 7630 & 7800\,$\pm$\,110\\
   191635 & V2109 Cyg& 0.01& 0.06 & 0.09 & 7130 & 3.47 &      &      & 7210 & 7200 & 7260\,$\pm$\,150\\
   192871 & V383 Vul& 0.03 & 0.02 & 0.02 & 7130 & 3.44 &      &      & 7215 & 6920 & 7300\,$\pm$\,140\\
   199908 & DQ Cep  & 0.00 & 0.01 & 0.00 & 7130 & 3.56 & 6940 & 3.74 & 7100 & 7030 & 7260\,$\pm$\,100\\
   210957 &         & 0.02 & 0.01 & 0.01 & 7870 & 3.80 &      &      & 7560 & 7670 & 7600\,$\pm$\,100\\
   213272 &         & 0.00 & 0.01 & 0.00 & 9100 & 4.07 & 9185 & 4.16 & 9000 &      & 9115\,$\pm$\,130\\
   214698 & 41 Peg  & 0.00 & 0.05 & 0.05 & 9113 & 3.69 & 9840 & 3.93 & 9480 &      & 9445\,$\pm$\,160\\
   219586 & V388 Cep&      & 0.01 & 0.00 &      &      &      &      & 7490 & 7235 & 7190\,$\pm$\,210\\
   034409 & BS Cam  & 0.09 & 0.10 & 0.06 & 6995 & 3.22 &      &      & 6950 & 6900 & 7360\,$\pm$\,270\\
   037819 & V356 Aur& 0.14 & 0.14 & 0.09 & 6780 & 3.78 & 6450 & 2.84 & 6660 & 6670 & 6800\,$\pm$\,220\\
   050018 & OX Aur  &0.02  & 0.01 & 0.00 & 6810 & 3.44 &      &      & 6870 & 6730 & 6840\,$\pm$\,145\\
   060302 & V344 Gem& 0.00 & 0.01 & 0.00 & 7310 & 3.78 &      &      & 7430 & 7170 & 7200\,$\pm$\,130\\
   079781 & GG UMa  & 0.00 & 0.09 & 0.06 & 6930 & 3.84 &      &      & 6900 & 6970 & 6930\,$\pm$\,200\\
   081882 & KZ UMa  & 0.00 & 0.05 & 0.01 & 7220 & 3.55 &      &      & 7270 & 7470 & 7810\,$\pm$\,130\\ 
   082620 & DL UMa  & 0.00 & 0.01 & 0.00 & 7175 & 3.74 &      &      & 7100 & 7265 & 7460\,$\pm$\,140\\
   084800 & IX UMa  &      & 0.02 & 0.00 &      &      &      &      & 8650 &      & 8760\,$\pm$\,190\\
   090747 & GS UMa  &      & 0.04 & 0.01 &      &      &      &      & 6550 & 6430 & 6610\,$\pm$\,120\\
   093044 & EO UMa  & 0.00 & 0.01 & 0.01 & 7230 & 3.83 & 7270 & 4.18 & 7380 & 7190 & 7380\,$\pm$\,240\\
   097302 & FI UMa  & 0.01 & 0.02 & 0.02 & 8500 & 4.19 & 8850 & 4.35 & 8755 &      & 8870\,$\pm$\,120\\
   099983 & HQ UMa  & 0.01 & 0.03 & 0.01 & 7145 & 3.52 &      &      & 7160 & 7030 & 7300\,$\pm$\,100\\
   102355 & KW UMa  & 0.00 & 0.03 & 0.01 & 7550 & 3.71 &      &      & 7710 & 7460 & 7900\,$\pm$\,160\\
   118954 & IP UMa  & 0.04 & 0.01 & 0.01 & 7510 & 3.79 &      &      & 7240 & 7200 & 7550\,$\pm$\,130\\
   127411 & IT Dra  &      & 0.00 & 0.01 &      &      &      &      & 8270 & 8100 & 7940\,$\pm$\,110\\
   151938 & V919 Her&      & 0.01 & 0.00 &      &      &      &      & 7185 & 7105 & 7015\,$\pm$\,100\\
   154225 & V929 Her& 0.00 & 0.01 & 0.01 & 6950 & 3.90 &      &      & 6480 & 6780 & 7130\,$\pm$\,250\\
   155118 & V873 Her& 0.06 & 0.02 & 0.05 & 7190 & 3.66 &      &      & 7160 & 7030 & 7120\,$\pm$\,200\\
   161287 & V966 Her&0.00  & 0.02 & 0.02 & 7110 & 3.68 &      &      & 7160 & 7170 & 7500\,$\pm$\,270\\
   176445 & V1438 Aql& 0.09& 0.02 & 0.07 & 7015 & 3.48 &      &      & 6850 & 6860 & 7150\,$\pm$\,120\\
   176503 & V544 Lyr& 0.06 & 0.02 & 0.02 & 7900 & 3.60 &      &      & 7360 & 7410 & 7700\,$\pm$\,160\\
   184522 & V2084 Cyg&     & 0.01 & 0.01 &      &      &      &      & 7030 & 7090 & 6895\,$\pm$\,100\\
   453111 & V456 Aur & 0.04& 0.02 & 0.03 & 6970 & 3.67 &      &      & 7030 & 6860 & 7150\,$\pm$\,200\\
\bottomrule
\end{tabular*}
\end{table*} 

\section{Atmospheric parameters from Photometry and Spectral energy distribution}

Initial atmospheric parameters, \teff\ and \logg\, were obtained from the photometric indices and the spectral energy distribution (SED). 
These parameters were used as inputs in the subsequent spectral analysis.

The interstellar reddening, ${\it E(B-V)}$, has a notable influence on the values of atmospheric parameters determined by photometric indices.
Hence, the ${\it E(B-V)}$ value should be taken into account in the analysis.
First, we derived ${\it E(B-V)}$ values using an interstellar extinction map and Na\,D$_{2}$ interstellar line. 
The ${\it E(B-V)}$ values were first determined from the interstellar extinction map \citep{2005AJ....130..659A} using the Galactic coordinates and distances of targets. 
We used the {\sc Hipparcos} parallaxes \citep{2007A&A...474..653V} for the stars which have distances less than $100$ parsec
and Gaia parallaxes \citep{2017A&A...599A..67C} for the stars which have distances greater than $100$ parsec.
The Gaia parallaxes are more accurate for the distances greater than $100$ parsec. 
For the member of Pleiades cluster V624\,Tau, we used the cluster distance to calculate the ${\it E(B-V)}$, because the parallax of this star has not been measured.  
The errors of ${\it E(B-V)}$ mainly come from the uncertainty of targets' distances. 
The largest error of ${\it E(B-V)}$ value was found to be $0.08$\,mag. 
In the second method, we calculated ${\it E(B-V)}$ values from ${uvby\beta}$ photometry. The ${\it E(b-y)}$ values 
were firstly calculated utilizing the method of \citet{1985MNRAS.217..305M}. Using the transformation 
${\it E(B-V)}$ $=$ 1.4 ${\it E(b-y)}$ \citep{1989ApJ...345..245C},
the ${\it E(B-V)}$ values were found and the average error of these ${\it E(B-V)}$ values was obtained to be $0.02$\,mag.  In the last method,
the ${\it E(B-V)}$ was derived by using the relation between equivalent width of Na\,D$_{2}$ line ($5889.95$\,{\AA})
and ${\it E(B-V)}$ \citep{1997A&A...318..269M}.
In this method uncertainties of the ${\it E(B-V)}$ values were adopted to be $0.02$\,mag \citep{2016MNRAS.458.2307K} (hereafter KA16). 
The obtained ${\it E(B-V)}$ values are listed in Table\,3.
As can be seen, ${\it E(B-V)}$ determined with all methods are generally in agreement with each other within error bars.
In the photometric analysis, the more accurate ${\it E(B-V)}$ values determined from the Na\,D$_{2}$ line were used.

In the next step, the atmospheric parameters, \teff\ and \logg, were derived using the de-reddened indices of
Johnson, 2MASS, uvby$\beta$\ Str\"{o}mgren and Geneva photometric systems. 
The photometric data were gathered from the General Catalogue of photometric data (GCPD\footnote{http://obswww.unige.ch/gcpd/gcpd.html}) \citep{1997A&AS..124..349M},
the 2MASS catalogue \citep{2003tmc..book.....C}, and the updated catalogue of Str\"{o}mgren\,-\,Crawford ${uvby\beta}$ 
photometry \citep{2015A&A...580A..23P}.
The \teff\ and \logg\ parameters were derived using the methods described by \citet{1985MNRAS.217..305M}, \citet{1997A&AS..122...51K}, \citet{2000AJ....120.1072S} 
and \citet{2006A&A...450..735M} for the Str\"{o}mgren, Geneva, Johnson and 2MASS systems, respectively. 
The \teff\ parameters were determined using all mentioned photometric systems,
while \logg\ parameters were obtained from Str\"{o}mgren and Geneva photometric systems only. 
In the calculations of \teff\ from Johnson, and 2MASS photometry, $\log g = 4.0$ and solar metallicity were assumed. 
The uncertainties of determined \teff\ and \logg\ parameters were calculated taking into account errors of ${\it E(B-V)}$ and used indices.
It turned out that the error on ${\it E(B-V)}$ contributes most to the uncertainty ($\sim 90$\,\%). 
The calculated atmospheric parameters and their uncertainties are given in Table\,3. 

In the last step, the \teff\ parameters were determined by using SED. 
In the analysis, the code written by Dr. Shulyak (private information) was used. 
The code automatically scans some spectrophotometric and photometric catalogues
and offers the possibility to manually add some photometric data (for more information see KA16).
In these calculations, solar metallicity and $\log g = 4.0$ were fixed. 
Final parameters are derived by comparing input data with the calculated theoretical spectra \citep[ATLAS9 code]{1993KurCD..13.....K}. 
The obtained \teff\ values and their errors are given in Table\,3.

\begin{figure*}
\includegraphics[width=17cm, angle=0]{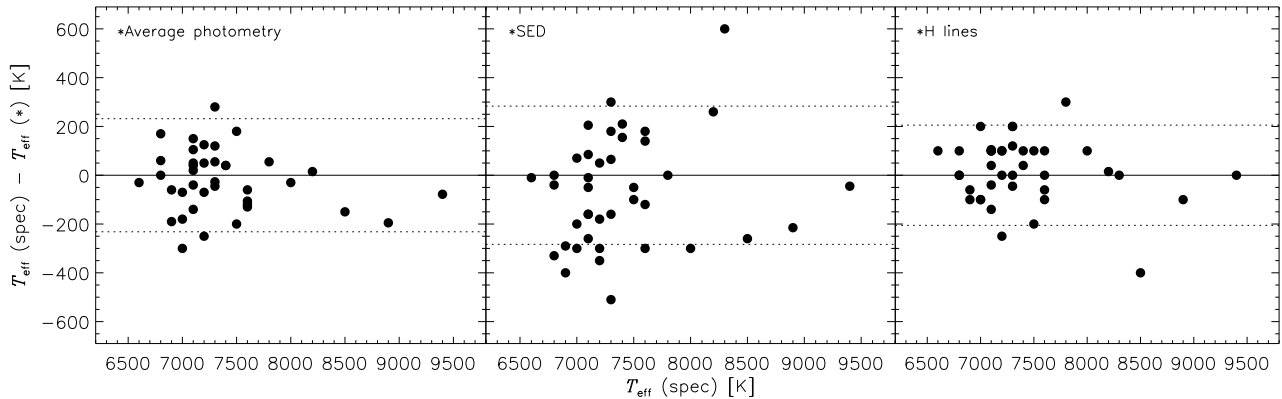}
\caption{Comparison of spectroscopic \teff\ values with the average photometric values, SED results and these from hydrogen lines analysis. 
The dotted lines represent 1-$\sigma$ level.}
\label{figure1}
\end{figure*}

\section{Spectroscopic atmospheric parameters and chemical abundance analysis}

Stellar atmospheric parameters have been determined from Balmer and metal lines analysis. 
Atmospheric chemical abundances were obtained using the spectral synthesis method. 
The hydrostatic, plane\,-\,parallel, and line\,-\,blanketed local thermodynamic equilibrium (LTE) 
atmosphere models were calculated with the ATLAS9 code \citep{1993KurCD..13.....K}, whereas 
the synthetic spectra were obtained with the SYNTHE code \citep{1981SAOSR.391.....K}.

\subsection{Analysis of Balmer lines}

To obtain \teff\ values, H$\beta$, H$\gamma$, and H$\delta$ lines were used. 
The HIDES spectra are in the range from $4080$ to $7525$\,{\AA}, hence the H$\delta$ line profiles were not used in their analysis. 
The procedure described by \citet{2004A&A...425..641C} was applied to derive \teff\ values from the analysis of Balmer lines. 
For stars with initial \teff\ lower than $8000$\,K, the \logg\ values were adopted to be $4.0$, 
as the Balmer lines are not sensitive to \logg\ for such temperatures \citep{2002A&A...395..601S, 2005MSAIS...8..130S}. 
For stars with estimated \teff\ values higher than $8000$\,K, both \teff\ and \logg\ were derived simultaneously. 
Additionally, metallicities of the stars were assumed to be $0.0$\,dex.
The analysis was performed separately for each Balmer line.
Final averaged \teff\ and \logg\ (for stars with \teff\ $> 8000$\,K) values are given in Table\,4. 

To estimate the uncertainties of \teff\ and \logg\ values determined from the Balmer lines analysis, 
we checked both, errors caused by the normalisation process and introduced by the assumed parameters such as \logg, metallicity, and \vsini.
The correct normalisation of Balmer lines is difficult, especially in the case of broad Balmer lines, which can be spread over more than one \'{e}chelle order. 
The error of \teff\ caused by inaccurate normalisation has been estimated as approximately $100$\,K
by checking the standard deviation of the different Balmer line's determinations.
When we took into account the effects of wrongly assumed values of \logg, metallicity, and \vsini, an average uncertainty of \teff\ was found to be $200$\,K. 
The uncertainties of \logg\ were obtained in a similar way, taking into account normalisation problems and effects of assumed parameters. 
The total uncertainties of \teff\ and \logg\ parameters were obtained using the squared sum of the individual contributions. 
The obtained parameters and their uncertainties are given in Table\,4.

\subsection{Analysis of metal lines}

The final atmospheric parameters, \teff, \logg, and microturbulent velocities ($\xi$) were obtained using the neutral and ionised iron lines. 
Previously determined atmospheric parameters were used as input values. The analysis was performed in the following steps:

\begin{description}

\item[--] The normalised spectra were divided into shorter spectral parts taking into account \vsini\ values of the stars.
For fast rotating objects (\vsini\,$>100$\,km\,s$^{-1}$) long parts covering many blended spectral lines were analysed, 
whereas for slowly rotating objects (\vsini\,$<30$\,km\,s$^{-1}$) short parts with one or few lines were taken into account. 

\item[--] The line identification for each part was performed using the line list of Kurucz \footnote{kurucz.harvard.edu/linelists.html}. 

\item[--] Analysis of the metal lines was performed for a range of \teff, \logg, and $\xi$ by using the spectrum synthesis method. 
The detailed information about the method can be found in \citet{2006ESASP.624E.120N}.
The analysis was carried out in steps of $100$\,K, $0.1$, and $0.1$\,km\,s$^{-1}$ for \teff, \logg, and $\xi$, respectively.

\item[--] The atmospheric parameters were obtained using the excitation potential (for \teff) and ionisation balance (for \logg)
of neutral and ionised iron lines. The $\xi$ parameters were derived checking the correlation between the iron abundances
and depths of iron lines (for more details see KA16). Simultaneously, \vsini\ parameters were adjusted.

\end{description}

\begin{figure*}
\includegraphics[width=17cm, angle=0]{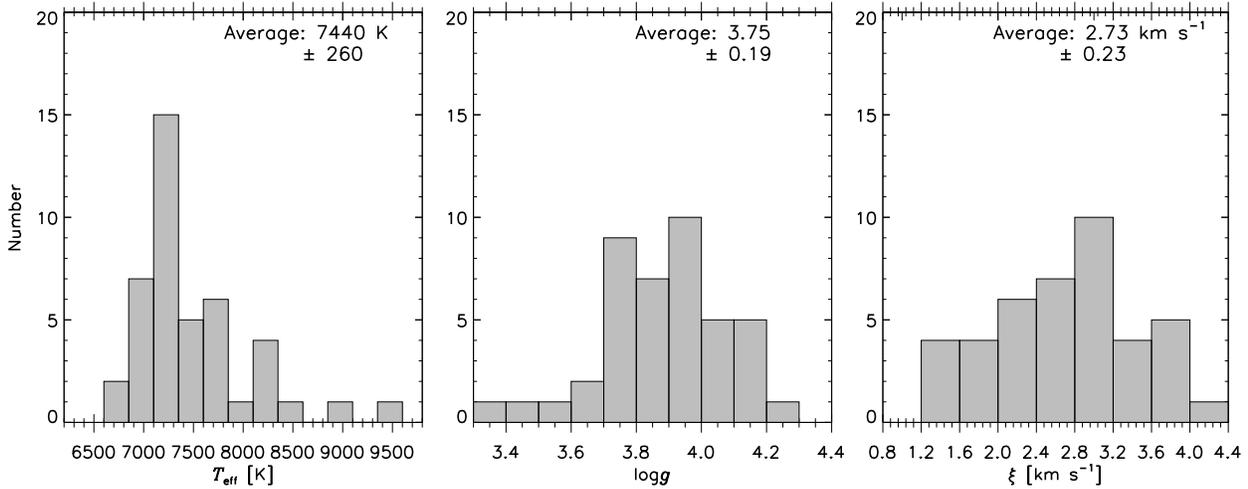}
\caption{Distributions of the atmospheric parameters of the analysed $\delta$\,Sct stars.}
\label{figure2}
\end{figure*}

The final atmospheric parameters are listed in Table\,4. 
The \teff\ values derived from different photometries and SEDs were compared with the spectroscopic \teff\ values in Fig.\,\ref{figure1}. 
As can be seen, these values are generally in agreement within 1-$\sigma$ level.
In Fig.\,\ref{figure2}, the distributions of the spectroscopic atmospheric parameters are presented. 

The uncertainties of the spectroscopic atmospheric parameters were determined considering relations between 
the iron abundances and both, excitation potentials of neutral or ionised iron lines, and lines depths.
For the accurate parameters, there is no correlation between them. It means that we have similar iron abundances, 
regardless of line excitation potential or line depth. To find the errors of \teff, \logg, and $\xi$ we check 
how these parameters change for correlation differences of about $5$\,\%. 
Using this method the maximum uncertainties of \teff, \logg, and $\xi$ were obtained to be $300$\,K, $0.3$\,dex, and $0.4$\,km\,s$^{-1}$, respectively.

\begin{figure}
\includegraphics[width=8.5cm, angle=0]{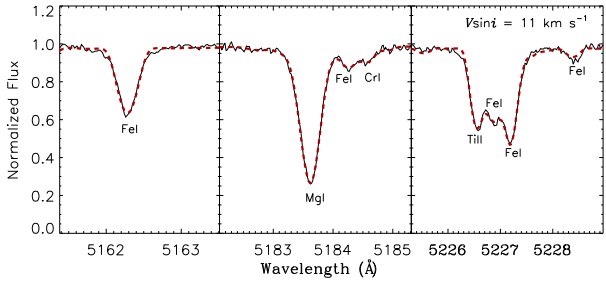}
\caption{Comparison of the theoretical (dashed line) and observed spectra (continuous line) of HD\,161287, one of the slowest rotating stars in our sample.}
\label{figure3}
\end{figure}

After the determination of accurate atmospheric parameters, the abundance analysis was performed with the spectrum synthesis method.
An example of fitting of the synthetic spectrum to the observed one is shown in Fig.\,\ref{figure3} for the slowly rotating star HD\,161287.

\begin{figure}
\includegraphics[width=8.5cm, angle=0]{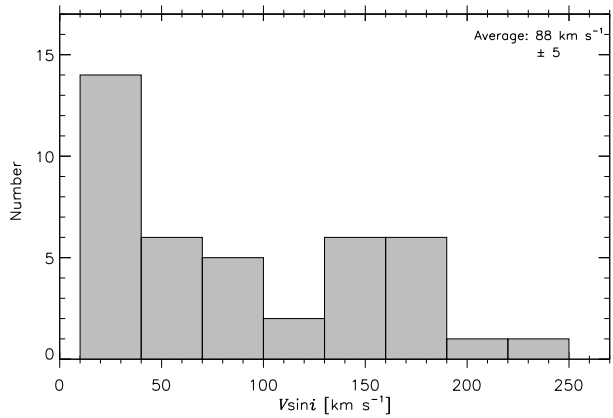}
\caption{Distribution of \vsini\ values of stars.}
\label{figure4}
\end{figure}

The \vsini\ values and the chemical abundances are shown in Table\,4 and Table\,5, respectively.
The distribution of the \vsini\ values is given in Fig.\,\ref{figure4}.
The obtained \vsini\ range from $10$ to $222$\,km\,s$^{-1}$.

In Table\,5 chemical abundances and their standard deviations are given for five analysed stars. The full table is given in the electronic form. 
The total errors of the determined abundances are due to the uncertainties of \teff, \logg,  $\xi$, and \vsini, 
assumptions adopted to calculate atmospheric models and synthetic spectra, quality of the data (resolving power, S/N), and the normalisation of spectra. 

The assumptions adopted in the atmospheric model calculations (e.g. LTE, plane-parallel geometry, and hydrostatic equilibrium)
cause an error of about $0.1$\,dex in chemical abundances \citep{2011mast.conf..314M}.
The uncertainties due to the resolving power and the S/N ratio of a spectrum were examined by KA16 and \citet{2016MNRAS.456.1221R}.
KA16 found uncertainties of about $0.07$\,dex and $0.13$\,dex for the iron abundance,
resulting from the resolving power difference (R\,=\,$67000$ and R\,=\,$80000$), and the S/N ratio difference (S/N\,=\,$310$ and S/N\,=\,$170$), respectively. 
In our study, such introduced uncertainties in chemical abundances cannot be checked, 
as we do not have even one star observed by different instruments. Therefore these uncertainties were adopted from KA16. 

To find the uncertainties of chemical abundances introduced by possible errors in atmospheric parameters \teff, \logg, and $\xi$,
we have selected a few stars with effective temperatures typical for the analysed sample.
We were changing values of their atmospheric parameters to check how such changes will influence the determined abundances of chemical elements.
If the error of \teff\ equals $100$\,K, the abundance of iron will change of $0.05$\,dex.
Smaller differences, $0.01$ and $0.02$\,dex, are caused by the 0.1 error of \logg\ and the $0.1$\,\kms uncertainty of $\xi$.
The \vsini\ effect on chemical abundances was examined as well. 
We found that the uncertainties in abundances caused by \vsini\ are in a range of $0.05-0.25$\,dex. 
The higher error values were obtained for stars with the higher \vsini\ values. 

The combined errors calculated taking into account all mentioned effects are given in Table\,4 for the iron abundances. 

\begin{table*}
\centering
  \caption{Atmospheric parameters derived from Balmer and iron line analysis.}
  \begin{tabular*}{0.9\linewidth}{@{\extracolsep{\fill}}rllllcrc}
\toprule
          &          &\multicolumn{1}{c}{\hrulefill Balmer lines\,\hrulefill}
                     &\multicolumn{5}{c}{\hrulefill \,Fe lines\,\hrulefill}\\
     HD   & Name     &\teff\           & \teff\           &  \logg\         &  $\xi$ & \vsini\       & $\log \epsilon$ (Fe) \\
   number &	         &    (K)          &     (K)	      &	                &  (\kms)&	(\kms)       &                      \\  \midrule
   
  089843 & EN UMa   & 7300\,$\pm$\,300 & 7400\,$\pm$\,300 & 4.0\,$\pm$\,0.2 & 4.0\,$\pm$\,0.4 & 175\,$\pm$\,10 & 7.30\,$\pm$\,0.68 \\
  099002 & CX UMa   & 7200\,$\pm$\,300 & 7200\,$\pm$\,200 & 4.2\,$\pm$\,0.2 & 4.0\,$\pm$\,0.3 & 145\,$\pm$\,6  & 7.38\,$\pm$\,0.62 \\
  115308 & DK Vir   & 6800\,$\pm$\,300 & 7100\,$\pm$\,200 & 4.0\,$\pm$\,0.2 & 2.7\,$\pm$\,0.2 & 75\,$\pm$\,3   & 7.66\,$\pm$\,0.56 \\
         & EH Lib   & 7100\,$\pm$\,200 & 7300\,$\pm$\,100 & 3.9\,$\pm$\,0.1 & 2.5\,$\pm$\,0.2 & 15\,$\pm$\,1   & 7.11\,$\pm$\,0.45 \\
  010088 &          & 7100\,$\pm$\,200 & 7300\,$\pm$\,200 & 4.1\,$\pm$\,0.2 & 2.2\,$\pm$\,0.2 & 45\,$\pm$\,2   & 7.93\,$\pm$\,0.57 \\
  012389 &          & 7900\,$\pm$\,300 & 8000\,$\pm$\,200 & 3.8\,$\pm$\,0.2 & 2.2\,$\pm$\,0.3 & 85\,$\pm$\,5   & 7.11\,$\pm$\,0.65 \\
  023156 & V624 Tau & 7600\,$\pm$\,200 & 7600\,$\pm$\,200 & 4.0\,$\pm$\,0.1 & 1.9\,$\pm$\,0.1 & 35\,$\pm$\,3   & 7.63\,$\pm$\,0.50\\
  062437 & AZ CMi   & 7500\,$\pm$\,200 & 7600\,$\pm$\,100 & 3.9\,$\pm$\,0.1 & 2.4\,$\pm$\,0.2 & 42\,$\pm$\,3   & 7.56\,$\pm$\,0.45\\
  073857 & VZ Cnc   & 7000\,$\pm$\,200 & 6900\,$\pm$\,200 & 3.4\,$\pm$\,0.2 & 2.5\,$\pm$\,0.1 & 26\,$\pm$\,1   & 7.34\,$\pm$\,0.54 \\
  103313 & IQ Vir   & 7700\,$\pm$\,200 & 7600\,$\pm$\,200 & 4.0\,$\pm$\,0.2 & 2.5\,$\pm$\,0.2 & 64\,$\pm$\,3   & 7.63\,$\pm$\,0.42\\
  110377 & GG Vir   & 7500\,$\pm$\,300 & 7800\,$\pm$\,200 & 3.8\,$\pm$\,0.2 & 3.3\,$\pm$\,0.3 & 173\,$\pm$\,5  & 7.18\,$\pm$\,0.63\\
  191635 & V2109 Cyg& 6800\,$\pm$\,200 & 7100\,$\pm$\,200 & 3.8\,$\pm$\,0.2 & 3.0\,$\pm$\,0.2 & 15\,$\pm$\,3   & 7.67\,$\pm$\,0.51\\
  192871 & V383 Vul & 7000\,$\pm$\,300 & 6900\,$\pm$\,100 & 3.8\,$\pm$\,0.2 & 4.0\,$\pm$\,0.3 & 148\,$\pm$\,5  & 7.23\,$\pm$\,0.36\\
  199908 & DQ Cep   & 7000\,$\pm$\,200 & 7100\,$\pm$\,200 & 3.8\,$\pm$\,0.1 & 2.7\,$\pm$\,0.1 & 64\,$\pm$\,3   & 7.82\,$\pm$\,0.41\\
  210957 &          & 7700\,$\pm$\,300 & 7500\,$\pm$\,200 & 3.8\,$\pm$\,0.1 & 2.9\,$\pm$\,0.2 & 77\,$\pm$\,3   & 7.25\,$\pm$\,0.48\\
  213272 &          & 9000\,$\pm$\,300$^{*}$ & 8900\,$\pm$\,200 & 3.9\,$\pm$\,0.2 & 2.1\,$\pm$\,0.3 & 133\,$\pm$\,3  & 7.09\,$\pm$\,0.55\\
  214698 & 41 Peg   & 9400\,$\pm$\,200$^{*}$ & 9400\,$\pm$\,200 & 3.7\,$\pm$\,0.2 & 1.2\,$\pm$\,0.3 & 34\,$\pm$\,2   & 7.70\,$\pm$\,0.40\\
  219586 & V388 Cep & 7300\,$\pm$\,300 & 7400\,$\pm$\,200 & 4.0\,$\pm$\,0.2 & 3.9\,$\pm$\,0.3 & 169\,$\pm$\,5  & 7.33\,$\pm$\,0.57\\
  034409 & BS Cam   & 7000\,$\pm$\,300 & 6900\,$\pm$\,200 & 3.5\,$\pm$\,0.3 & 2.5\,$\pm$\,0.3 & 197\,$\pm$\,10 &7.81\,$\pm$\,0.74\\
  037819 & V356 Aur & 6700\,$\pm$\,300 & 6900\,$\pm$\,200 & 4.0\,$\pm$\,0.2 & 3.1\,$\pm$\,0.2 & 24\,$\pm$\,2   & 7.89\,$\pm$\,0.48\\
  050018 & OX Aur   & 6800\,$\pm$\,300 & 6800\,$\pm$\,300 & 3.6\,$\pm$\,0.2 & 4.1\,$\pm$\,0.3 & 155\,$\pm$\,12 & 7.33\,$\pm$\,0.68\\
  060302 & V344 Gem & 7000\,$\pm$\,300 & 7300\,$\pm$\,200 & 4.0\,$\pm$\,0.2 & 2.2\,$\pm$\,0.3 & 143\,$\pm$\,5  & 7.51\,$\pm$\,0.69\\
  079781 & GG UMa   & 6700\,$\pm$\,300 & 6900\,$\pm$\,200 & 4.1\,$\pm$\,0.2 & 2.9\,$\pm$\,0.3 & 64\,$\pm$\,3   & 7.59\,$\pm$\,0.56\\
  081882 & KZ UMa   & 7400\,$\pm$\,300 & 7200\,$\pm$\,300 & 3.7\,$\pm$\,0.3 & 3.7\,$\pm$\,0.2 & 117\,$\pm$\,5  & 7.27\,$\pm$\,0.53\\
  082620 & DL UMa   & 7200\,$\pm$\,300 & 7400\,$\pm$\,200 & 4.2\,$\pm$\,0.1 & 3.6\,$\pm$\,0.2 & 68\,$\pm$\,2   & 7.44\,$\pm$\,0.54\\
  084800 & IX UMa   & 8200\,$\pm$\,300$^{*}$ & 8400\,$\pm$\,200 & 4.2\,$\pm$\,0.2 & 1.3\,$\pm$\,0.3 & 167\,$\pm$\,7  & 7.60\,$\pm$\,0.62\\
  090747 & GS UMa   & 6300\,$\pm$\,200 & 6600\,$\pm$\,300 & 4.0\,$\pm$\,0.2 & 1.5\,$\pm$\,0.2 & 34\,$\pm$\,2   & 7.41\,$\pm$\,0.57\\
  093044 & EO UMa   & 7200\,$\pm$\,300 & 7100\,$\pm$\,200 & 3.9\,$\pm$\,0.2 & 2.9\,$\pm$\,0.3 & 113\,$\pm$\,4  & 7.15\,$\pm$\,0.60\\
  097302 & FI UMa   & 7600\,$\pm$\,300 & 8100\,$\pm$\,300 & 4.2\,$\pm$\,0.2 & 3.3\,$\pm$\,0.3 & 132\,$\pm$\,6  & 7.69\,$\pm$\,0.75\\
  099983 & HQ UMa   & 6800\,$\pm$\,300$^{*}$ & 7000\,$\pm$\,200 & 4.3\,$\pm$\,0.2 & 1.2\,$\pm$\,0.2 & 161\,$\pm$\,8 & 7.84\,$\pm$\,0.64\\
  102355 & KW UMa   & 8000\,$\pm$\,300 & 7600\,$\pm$\,300 & 3.8\,$\pm$\,0.3 & 2.0\,$\pm$\,0.4 & 175\,$\pm$\,17 &7.22\,$\pm$\,0.77 \\
  118954 & IP UMa   & 7000\,$\pm$\,200 & 7500\,$\pm$\,200 & 3.9\,$\pm$\,0.1 & 3.1\,$\pm$\,0.2 & 26\,$\pm$\,2  & 7.64\,$\pm$\,0.48\\
  127411 & IT Dra   & 8100\,$\pm$\,300$^{*}$ & 8100\,$\pm$\,200 & 4.1\,$\pm$\,0.2 & 2.0\,$\pm$\,0.4 & 222\,$\pm$\,9& 7.19\,$\pm$\,0.72\\
  151938 & V919 Her & 7000\,$\pm$\,200 & 7200\,$\pm$\,100 & 3.8\,$\pm$\,0.2 & 1.7\,$\pm$\,0.2 & 10\,$\pm$\,1 & 7.72\,$\pm$\,0.44\\
  154225 & V929 Her & 6800\,$\pm$\,200 & 6900\,$\pm$\,200 & 3.9\,$\pm$\,0.2 & 2.9\,$\pm$\,0.2 & 35\,$\pm$\,2  & 7.61\,$\pm$\,0.49\\
  155118 & V873 Her & 7100\,$\pm$\,300 & 7300\,$\pm$\,200 & 3.9\,$\pm$\,0.2 & 2.3\,$\pm$\,0.2 & 70\,$\pm$\,3 & 7.76\,$\pm$\,0.51\\
  161287 & V966 Her & 7000\,$\pm$\,200 & 7100\,$\pm$\,200 & 4.0\,$\pm$\,0.1 & 3.1\,$\pm$\,0.1 & 11\,$\pm$\,1 & 7.31\,$\pm$\,0.48\\
  176445 & V1438 Aql& 7100\,$\pm$\,300 & 7200\,$\pm$\,200 & 4.0\,$\pm$\,0.2 & 2.9\,$\pm$\,0.2 & 94\,$\pm$\,5 & 7.54\,$\pm$\,0.62\\
  176503 & V544 Lyr & 7600\,$\pm$\,200 & 8100\,$\pm$\,200 & 3.8\,$\pm$\,0.2 & 2.8\,$\pm$\,0.2 & 10\,$\pm$\,1 &7.94\,$\pm$\,0.50\\
  184522 & V2084 Cyg& 6700\,$\pm$\,200 & 7100\,$\pm$\,200 & 4.1\,$\pm$\,0.2 & 3.2\,$\pm$\,0.2 & 37\,$\pm$\,3 & 7.51\,$\pm$\,0.51\\
  453111 & V456 Aur & 7000\,$\pm$\,200 & 7100\,$\pm$\,200 & 4.1\,$\pm$\,0.1 & 3.3\,$\pm$\,0.1 & 24\,$\pm$\,1 & 7.22\,$\pm$\,0.49\\
\bottomrule
\end{tabular*}
\begin{description}
 \item[]$^{*}$\logg\ values were determined from Balmer lines' analysis: HD\,213272: \logg\,$=4.0\pm0.1$, HD\,214698: \logg\,$=3.7\pm0.1$, HD\,84800: \logg\,$=3.8\pm0.2$, HD\,102355: \logg\,$=3.6\pm0.1$, HD\,127411: \logg\,$=4.0\pm0.3$.  
\end{description}
\end{table*}

\begin{table*}
\centering
  \caption{Average abundances and standard deviations of individual elements. Number of the analysed spectral parts is given in the brackets. The full table is available in the electronic form.}
  \begin{tabular*}{0.9\linewidth}{@{\extracolsep{\fill}}llllll}
\toprule
  Elements & HD\,89843        & HD\,99002             & HD\,115308            & HD\,10088             &HD\,23156		  \\ 
(atomic number)               &                       &                       &                       & 			      \\
\midrule
C  (6) &8.99\,$\pm$\,0.21 (2) & 8.71\,$\pm$\,0.20 (2) &8.83\,$\pm$\,0.24 (4)  &7.92\,$\pm$\,0.32 (3)  &8.156\,$\pm$\,0.15 (4) \\
N (7)  &                      &		                  &                       &		                  &			   \\
O (8)  &9.21\,$\pm$\,0.21 (1) &   		              &8.97\,$\pm$\,0.24  (1) &8.46\,$\pm$\,0.32 (1)  &8.97\,$\pm$\,0.22 (1)  \\
Na (11)&                      &                       &5.74\,$\pm$\,0.24 (1)  &7.18\,$\pm$\,0.32 (2)  &6.44\,$\pm$\,0.22 (2)  \\
Mg (12)&7.85\,$\pm$\,0.21 (2) &                       &7.81\,$\pm$\,0.29 (5)  &7.64\,$\pm$\,0.14 (4)  &8.10\,$\pm$\,0.31 (5)  \\
Si (14)&6.90\,$\pm$\,0.21 (2) &                       &6.93\,$\pm$\,0.37 (9)  &7.12\,$\pm$\,0.52 (9)  &7.21\,$\pm$\,0.34 (10) \\
S (16) &7.63\,$\pm$\,0.21 (2) &		                  &7.17\,$\pm$\,0.24 (2)  &7.69\,$\pm$\,0.32 (1)  &7.43\,$\pm$\,0.22 (2)  \\
Ca (20)&6.41\,$\pm$\,0.34 (6) &6.03\,$\pm$\,0.31 (8)  &6.50\,$\pm$\,0.19 (15) &5.71\,$\pm$\,0.26 (12) &6.62\,$\pm$\,0.19 (13) \\
Sc (21)&2.49\,$\pm$\,0.21 (2) &3.28\,$\pm$\,0.11 (3)  &3.17\,$\pm$\,0.17 (6)  &1.95\,$\pm$\,0.32 (3)  &3.22\,$\pm$\,0.15 (8)  \\
Ti (22)&5.07\,$\pm$\,0.15 (7) &5.20\,$\pm$\,0.25 (15) &5.08\,$\pm$\,0.20 (30) &5.21\,$\pm$\,0.21 (23) &5.11\,$\pm$\,0.26 (23) \\
V (23) &4.50\,$\pm$\,0.21 (1) &		                  &4.42\,$\pm$\,0.24 (2)  &4.55\,$\pm$\,0.32 (2)  &4.239\,$\pm$\,0.22 (2) \\
Cr (24)&5.45\,$\pm$\,0.25 (7) &5.62\,$\pm$\,0.16 (9)  &5.38\,$\pm$\,0.21 (18) &6.09\,$\pm$\,0.18 (26) &5.82\,$\pm$\,0.31 (39) \\
Mn (25)&5.49\,$\pm$\,0.21 (1) &5.56\,$\pm$\,0.20 (3)  &4.90\,$\pm$\,0.31 (25) &5.59\,$\pm$\,0.21 (11) &5.25\,$\pm$\,0.22 (5)  \\
Fe (26)&7.30\,$\pm$\,0.17 (18)&7.38\,$\pm$\,0.17 (26) &7.25\,$\pm$\,0.15 (67) &7.93\,$\pm$\,0.21 (118)&7.63\,$\pm$\,0.17 (117)\\
Co (27)&                      &		                  &                       &		                  &                       \\
Ni (28)&6.33\,$\pm$\,0.14 (4) &6.14\,$\pm$\,0.22 (9)  &5.97\,$\pm$\,0.16 (16) &6.94\,$\pm$\,0.20 (16) &6.27\,$\pm$\,0.21 (16) \\
Cu (29)&                      &		                  &                       &		                  &4.82\,$\pm$\,0.22 (1)  \\
Zn (30)&                      &4.42\,$\pm$\,0.20 (1)  &		                  &4.86\,$\pm$\,0.32 (2)  &4.10\,$\pm$\,0.22 (1)  \\
Sr (38)&                      &		                  &3.21\,$\pm$\,0.24 (2)  &3.80\,$\pm$\,0.32 (2)  &3.72\,$\pm$\,0.22 (2)  \\
Y (39) &3.21\,$\pm$\,0.21 (2) &2.90\,$\pm$\,0.17 (4)  &2.25\,$\pm$\,0.43 (3)  &3.58\,$\pm$\,0.32 (2)  &2.59\,$\pm$\,0.22 (2)  \\
Zr (40)&3.33\,$\pm$\,0.21 (1) &3.16\,$\pm$\,0.20 (1)  &2.63\,$\pm$\,0.24 (3)  &3.00\,$\pm$\,0.32 (2)  &2.98\,$\pm$\,0.22 (2)  \\
Ba (56)&1.94\,$\pm$\,0.21 (2) &                       &2.85\,$\pm$\,0.24 (2)  &3.82\,$\pm$\,0.32 (2)  &2.11\,$\pm$\,0.22 (2)  \\
\bottomrule
\end{tabular*}
\end{table*}

\section{Discussion of the results}

\subsection{Atmospheric parameters of $\delta$\,Sct stars}

The \teff\ range of investigated $\delta$\,Sct stars was found to be $6600-9400$\,K. 
The typical \teff\ values of $\delta$\,Sct stars vary from $6300$ to $8600$\,K \citep{2011A&A...534A.125U}. 
As can be seen, the derived \teff\ range is in agreement with the characteristic \teff\ values of $\delta$\,Sct stars. 
However, there are two hot stars (HD\,213272 and HD\,214698) that are located beyond the \teff\ range.
These stars are suspected $\delta$\,Sct variables (see Table\,2) and the nature of their variability should be checked.

We compared the determined range and average values of \teff\ for $\delta$\,Sct stars with those for $\gamma$\,Dor's obtained by KA16. 
The average \teff\ ($\sim 7440$\,$\pm$\,260\,K) of $\delta$\,Sct stars is slightly higher than 
this calculated for $\gamma$\,Dor's ($7060$\,$\pm$\,130\,K), as expected.
However, \teff\ ranges of both variables overlap.
This gives us an opportunity to check whether there is a chemical difference between both types of pulsating stars located 
in the same area of H-R diagram, close to the blue edge of the $\gamma$\,Dor instability strip (see Sect.\,6.4). 

The distribution of \logg\ determined from the iron lines analysis is shown in the middle panel of Fig.\,\ref{figure2}.
The \logg\ values were found between $3.4$ and $4.3$.
The obtained \logg\ values are in good agreement with the luminosity range (V-III) of $\delta$\,Sct stars.
Previous studies of $\delta$\,Sct stars \citep[e.g.][]{2008A&A...485..257F, 2011MNRAS.411.1167C} 
give \logg\ between $3.0$ and $4.3$ which are also in agreement with our results.
For $\gamma$\,Dor stars \logg\ ranges from $3.8$ to $4.5$ (KA16). 
It seems that $\delta$\,Sct stars are more evolved than $\gamma$\,Dor stars, as expected. 

\begin{figure}
\includegraphics[width=8.5cm, angle=0]{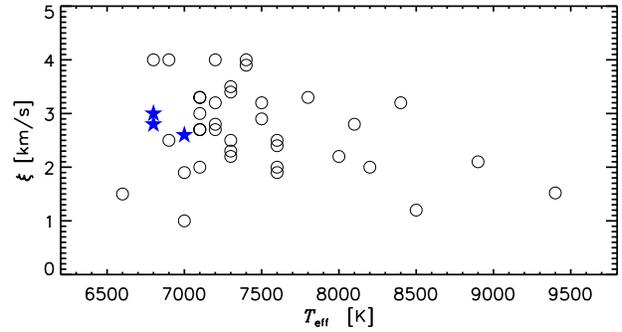}
\caption{$\xi$ as a function of \teff. The star symbol represents the chemically peculiar objects.}
\label{figure7}
\end{figure}

The distribution of the derived $\xi$ is shown in the right-hand panel of Fig.\,\ref{figure2}. 
The $\xi$ ranges from $1.2$ to $4.0$\,\kms\ with the average value $2.73$\,$\pm$\,0.23\,\kms. 
The $\xi$ values for $\gamma$\,Dor stars were found in the range from $1.3$ to $3.2$\,\kms with the average value $2.25$\,$\pm$\,0.2\,\kms\ (KA16).
The $\xi$ values of $\delta$\,Sct stars are higher than determined for $\gamma$\,Dor stars within errors.
It is consistent with the relation between the $\xi$ and \teff\ \citep[see e.g.][and references therein]{2015MNRAS.450.2764N}.
This relation was examined by \citet{2009A&A...503..973L}, \citet{2014psce.conf..193G}, \citet{2015MNRAS.450.2764N} and KA16.
It turned out that the $\xi$ value is inversely proportional to \teff\ for \teff\ higher than $\sim$7400\,K.
This relation for our stars is shown in Fig.\,\ref{figure7}. 
As can be seen from the figure, there is no difference between microturbulence values for CP and normal stars. 
Similar results were found by \citet{2015MNRAS.450.2764N} and KA16.

The \vsini\ values of the analysed stars range from $10$ to $222$\,\kms, with the average value $88$\,$\pm$\,5\,\kms. 
According to the previous studies \citep{2008A&A...485..257F, 2009LNP...765..207R, 2011MNRAS.411.1167C, 2013AJ....145..132C,2015MNRAS.450.2764N}, 
the \vsini\ values of A-F stars are in a range from $\sim 4$ to $300$\,\kms.
Our results are in agreement with the \vsini\ distributions given in the literature.
The \vsini\ values of $\gamma$\,Dor stars are from $5$ to $240$\,\kms\ and the average value is $81$\,$\pm$\,3\,\kms\ (KA16), 
similar to these obtained for $\delta$\,Sct stars.

\subsection{Correlations between the atmospheric and pulsation parameters of $\delta$\,Sct stars}

The possible correlations between the pulsation quantities (pulsation period $P_{\rm puls}$ and amplitude $Amp$) and the obtained parameters were examined.
For this purpose, the $P_{\rm puls}$ of the highest $Amp$ and these $Amp$ values in $V$-band were taken from \citet{2000A&AS..144..469R}.
Four high-amplitude $\delta$\,Sct stars (HADS)\footnote{The stars are signed in Table\,2.} are available in our sample. 
These stars were discarded in the following analysis. All the other objects represent typical $Amp$ values of classical $\delta$\,Sct stars. 

\begin{figure}
\includegraphics[width=8.5cm, angle=0]{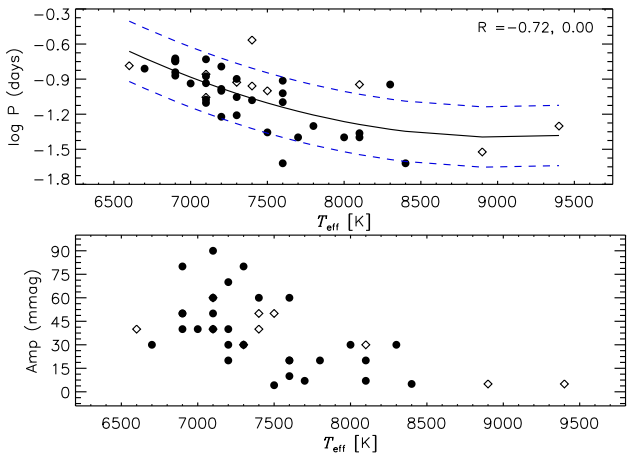}
\caption{Correlations between \teff\ and pulsation quantities. The solid and dashed lines in the upper panel show the correlation and 1-$\sigma$ level.
Circles and diamonds represent known and suspected $\delta$\,Sct stars, respectively. R is sperman rank and the first number in R shows the direction and strength of the correlation,
while the secondnumber represents deviations of points from the correlations.}
\label{figure6}
\end{figure}

In Fig.\,\ref{figure6}, the relations between \teff\ and pulsation quantities are shown.
As can be seen, there is an obvious correlation between \teff\ and $P_{\rm puls}$ and $Amp$.
These pulsation parameters have lower values for the hotter $\delta$\,Sct stars.
As known, changes in \teff\ are related to the stellar radii ($R$) and the changes in $R$ determine the position 
of helium ionisation zone which drives the $\delta$\,Sct type pulsations \citep{1980cox}.
Additionally, this negative correlation between $P_{\rm puls}$ and \teff\ can be explained taking into account basic equations.
Using the luminosity ($L$) -- mass ($M$) relation ($L/L_{\sun} \approx M/M_{\sun}$), the mean density ($\overline{\rho} \sim M/R^{3}$),
and the pulsation constant ($Q = P_{\rm puls} (\overline{\rho} / \overline{\rho}_{\sun})^{0.5}$), we can obtain that 
$P_{\rm puls} \varpropto (R/R_{\sun})^{0.5}$ (\teff\,/\,\teff$_{\sun}$)$^{-2}$.
\citet{1990DSSN....2...13B} also showed that the $P_{\rm puls}$ depends on \logg, \teff, and the bolometric magnitudes of pulsation stars. 
\citet{2011MNRAS.417..591B} found similar relationships between $P_{\rm puls}$, $Amp$ and \teff\ for $\delta$\,Sct stars in the {\it Kepler} field.
The \teff\,$-$\,$P_{\rm puls}$ relation was also found for $\delta$\,Sct stars in eclipsing binaries (Kahraman Ali\c{c}vu\c{s} et al., in preparation).
The same relations for $\gamma$\,Dor stars were checked by KA16. 
They did not find any significant correlation between \teff\ and $Amp$, while a weak and negative correlation was found between \teff\ and $P_{\rm puls}$.
This weak correlation is probably caused by the narrow \teff\ range of the analysed $\gamma$\,Dor stars which were used to check relations (from $6900$ to $7300$\,K).

\begin{figure}
\includegraphics[width=8.5cm, angle=0]{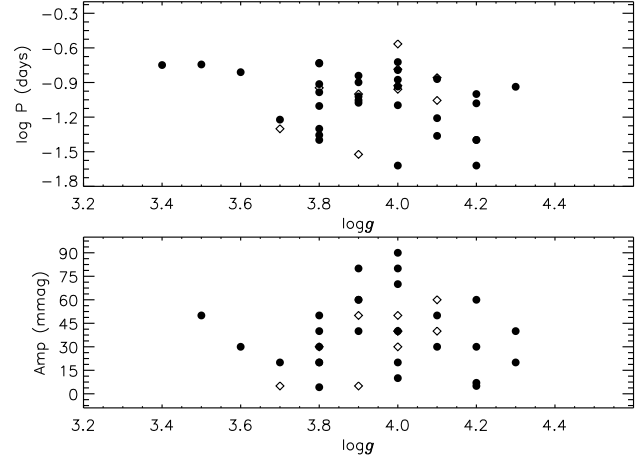}
\caption{The relations between pulsation quantities and \logg. Circles and diamonds represent known and suspected $\delta$\,Sct stars, respectively.}
\label{figure5_j}
\end{figure}

The correlations of \logg\ values with the pulsation quantities were examined as well.
The suitable relations are shown in the Fig.\,\ref{figure5_j}.
As can be seen, there are no significant correlations between \logg, $P_{\rm puls}$ and $Amp$. 
On the other hand, \citet{1990ASPC...11..481C} and \citet{1990DSSN....2...13B} found a relation between $P_{\rm puls}$ and \logg.
According to them, $P_{\rm puls}$ decreases with increasing \logg\ values. 
$\delta$\,Sct stars in our study were selected considering their positions in the \logg\ -- $\log$\,\teff\ diagram.
Most stars in our sample have \logg\ values from $3.8$ to $4.2$.
Only three stars have \logg\ below $3.6$ and the $P_{\rm puls}$ values for them are clearly higher than
the average period of stars with higher \logg\ values. 
So, the chosen sample of stars can be the reason why the \logg\,$-$\,$P_{\rm puls}$ is not observed.
The same relation was examined for $\gamma$\,Dor stars by KA16 and a weak and negative correlation was found between \logg\ and $Amp$,
while no correlation was found between \logg\ and $P_{\rm puls}$. 

\begin{figure}
\includegraphics[width=8.5cm, angle=0]{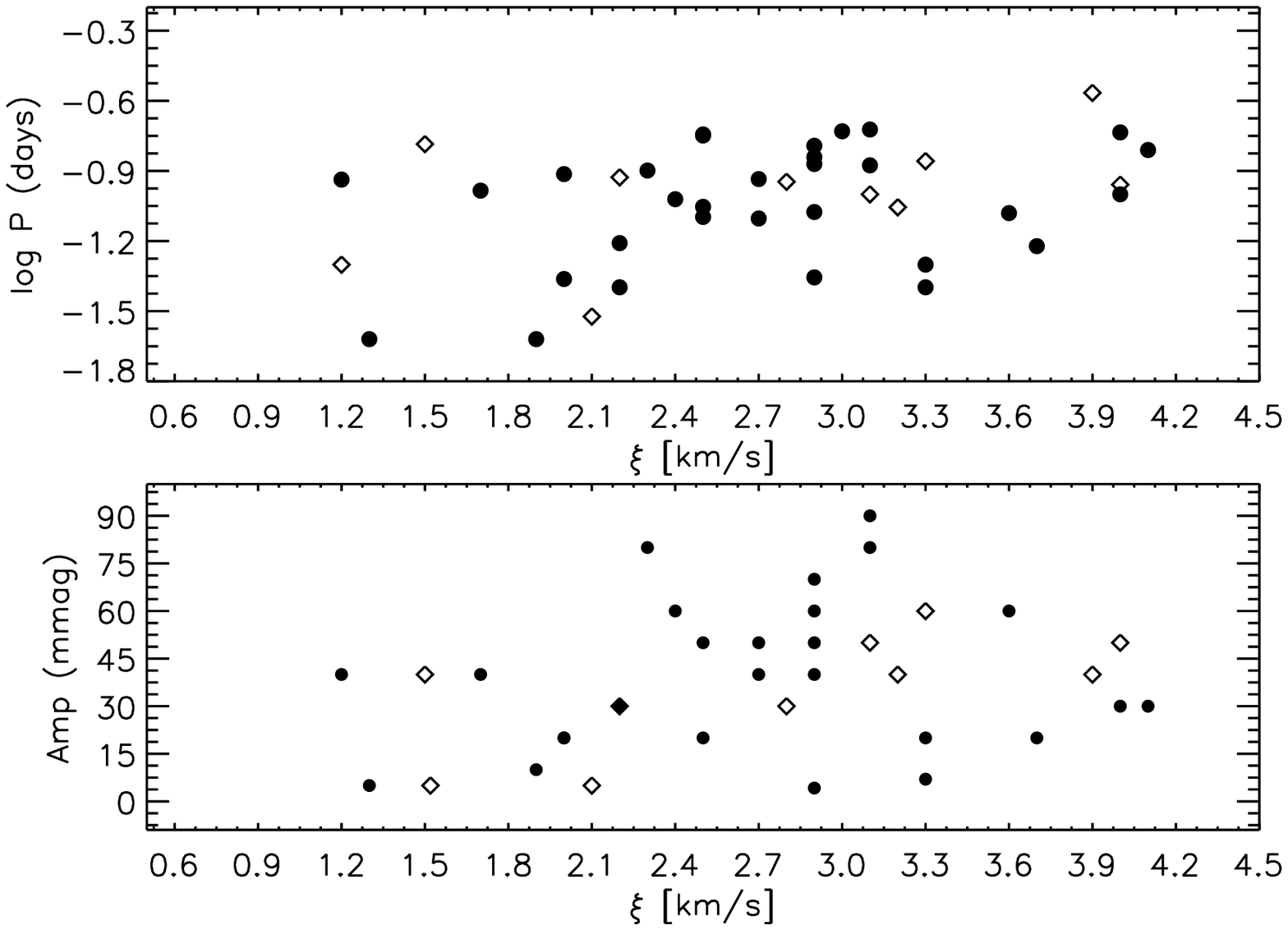}
\caption{The relations between the pulsation quantities and $\xi$. Circles and diamonds represent known and suspected $\delta$\,Sct stars, respectively.}
\label{figure8}
\end{figure}

Next, the possible correlations of $\xi$ values with the pulsation quantities were checked.
These relations are demonstrated in Fig.\,\ref{figure8}.
As can be seen, there are no significant correlations between $\xi$ and pulsation quantities. 
In general, higher values of $P_{\rm puls}$ correspond to higher values of $\xi$. 
The same correlation were examined in KA16 for $\gamma$\,Dor stars.
They found that a positive relation between $P_{\rm puls}$ and $\xi$ can occur. 

\begin{figure}
\includegraphics[width=8.5cm, angle=0]{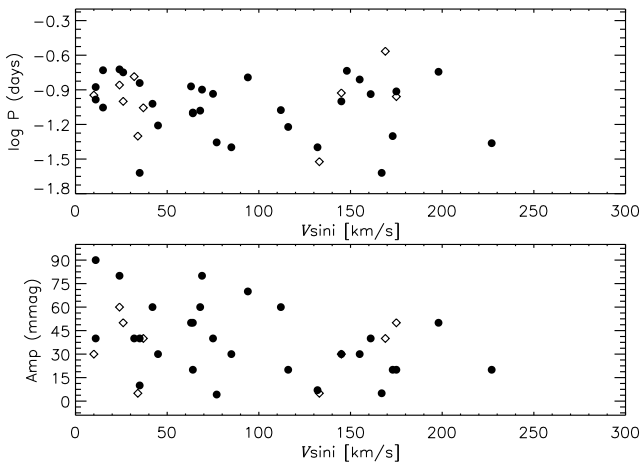}
\caption{The relations between the pulsation quantities and \vsini. Circles and diamonds represent known and suspected $\delta$\,Sct stars, respectively.}
\label{figure9}
\end{figure}

The relations between \vsini\ and pulsation quantities are shown in Fig.\,\ref{figure9}.
As can be seen, there are no significant correlations. 
A weak negative correlation may exist for \vsini\ and $Amp$.
Similar relations between \vsini\ and $P_{\rm puls}$ were also found in the 
literature \citep[e.g.][]{2000A&AS..144..469R, 2000ASPC..210....3B, 2013MNRAS.431.3685T}. 
For $\gamma$\,Dor stars a strong correlation between \vsini\ and $P_{\rm puls}$ and a weak correlation 
between \vsini\ and $Amp$ were found as well (\citet{2015ApJS..218...27V}, KA16). 

\begin{figure}
\includegraphics[width=8.5cm, angle=0]{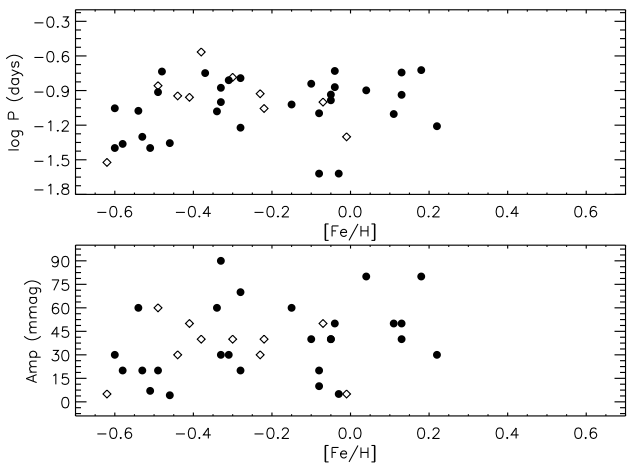}
\caption{The relations between pulsation quantities and metallicity. Circles and diamonds represent known and suspected $\delta$\,Sct stars, respectively.}
\label{figure13}
\end{figure}

Relations between the metallicity and pulsation quantities are presented in Fig.\,\ref{figure13}.
No correlations were found. However, the stars generally have metallicities lower than the solar one.
There is no significant difference between the average iron abundance of $\delta$\,Sct ($7.47$\,dex) and $\gamma$\,Dor ($7.41$\,dex) stars (KA16). 

\subsection{Positions of $\delta$\,Sct stars in the $\log$\,\teff\,$-$\,\logg\ diagram}

Positions of $\delta$\,Sct stars in the $\log$\,\teff\,$-$\,\logg\ diagram have been examined in detail
since new $\delta$\,Sct stars have been discovered by the space telescopes ({\it MOST} \citeauthor{2003PASP..115.1023W} \citeyear{2003PASP..115.1023W}, 
{\it CoRoT} \citeauthor{2009A&A...506..411A} \citeyear{2009A&A...506..411A}, {\it Kepler} \citeauthor{2010Sci...327..977B} \citeyear{2010Sci...327..977B}). 
\citet{2011A&A...534A.125U} showed positions of these variables in the $\log$\,\teff\,$-$\,\logg\ diagram mainly based on
the photometric atmosphere parameters from the KIC catalogue \citep{2011AJ....142..112B}. 
They found that {\it Kepler} $\delta$\,Sct stars are located both in their instability strip and outside of it.
In the most recent study \citep{2015MNRAS.450.2764N} accurate parameters of a few $\delta$\,Sct stars were obtained and 
some of them were found outside of their domain. However, the theory cannot explain this.

\begin{figure*}
\includegraphics[width=15cm, height=7cm, angle=0]{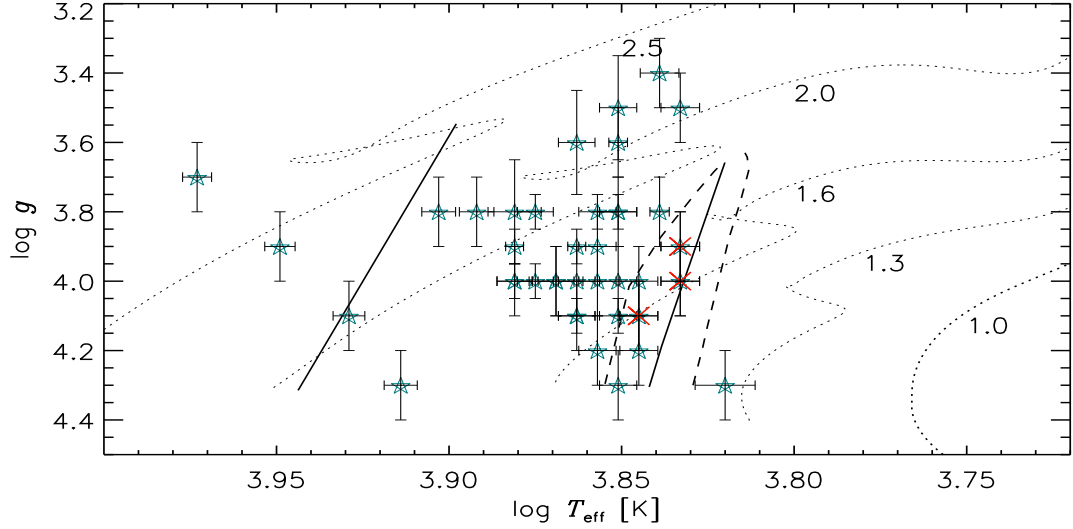}
\caption{Positions of $\delta$\,Sct stars in the theoretical instability strips of $\delta$\,Sct (solid lines) and $\gamma$\,Dor stars (dashed-lines) \citep{2005A&A...435..927D}.
Chemically peculiar stars are shown as crosses.}
\label{figure10}
\end{figure*}

In Fig.\,\ref{figure10} we show the positions of the analysed $\delta$\,Sct stars in the $\log$\,\teff\,$-$\,\logg\ diagram. 
As can be seen, most stars are located inside the $\delta$\,Sct instability strip.
However, there are a few stars placed outside of the $\delta$\,Sct instability strip.
HD\,213272 and HD\,214698 are located beyond the blue edge, while HD\,90747 is located outside the red edge.
The stars beyond the blue edge have the lowest amplitudes ($\sim 5$\,mmag), while the star on the cold side of the $\delta$\,Sct instability strip
does not show differences in pulsation quantities in comparison with the other $\delta$\,Sct stars considered. 
However, all these stars are suspected $\delta$ Sct variables (see Table\,2) and the verification of their variability types is necessary.

The positions of CP stars in the $\log$\,\teff\,$-$\,\logg\ diagram were also shown.
In a recent study, metallic stars with $\delta$\,Sct pulsations were found in the \teff\ range of $6900-7600$\,K \citep{2017MNRAS.465.2662S}. 
\teff\ of CP stars analysed here are also in this range (within error) and are located in the $\gamma$\,Dor instability strip.

\begin{figure}
\includegraphics[width=8.5cm, angle=0]{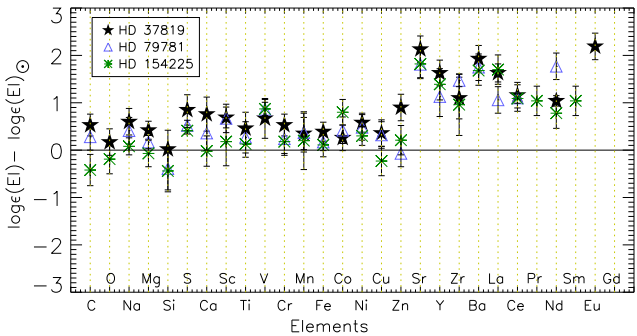}
\caption{Chemical abundances of chemically peculiar stars compared with the solar values \citep{2009ARA&A..47..481A}.}
\label{figure11}
\end{figure}

\subsection{Chemical abundances of $\delta$\,Sct stars}

During the spectral classification and abundance analysis, a few chemically peculiar stars were found.
HD\,37819, HD\,79781 and HD\,154225 have overabundant iron-peak and heavy elements (\ion{Zn}, \ion{Sr}, \ion{Zr}, and \ion{Ba}). 
The chemical abundance patterns of the discovered chemically peculiar stars are given in Fig\,\ref{figure11}.

\begin{figure}
\includegraphics[width=8.5cm, angle=0]{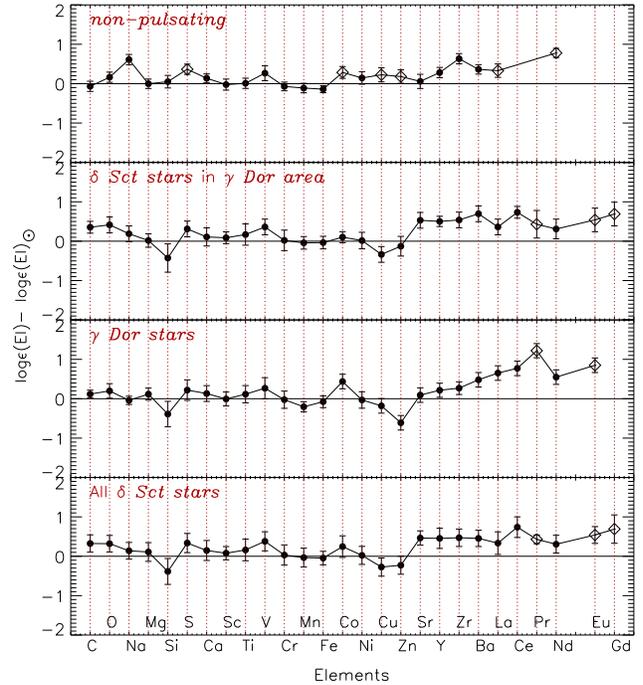}
\caption{Comparisons of the chemical abundance patterns of $\delta$\,Sct, $\gamma$\,Dor, and non-pulsating stars.
The chemical abundances of $\gamma$\,Dor, non-pulsating stars and solar abundances were taken from 
\citet{2016MNRAS.458.2307K, 2015MNRAS.450.2764N} and \citet{2009ARA&A..47..481A}, respectively. 
Filled circles and diamonds show values determined for more than $5$ stars or less than $5$ stars, respectively.}
\label{figure12}
\end{figure}

We compared the chemical abundance pattern of $\delta$\,Sct stars with the non-pulsating and $\gamma$\,Dor stars' patterns.
Atmospheric parameters and chemical abundances of eighteen non-pulsating stars were taken from \citet{2015MNRAS.450.2764N}. 
The stars used in the comparisons were analysed using exactly the same as in the present work. 
We divided $\delta$\,Sct stars into two groups: one contains all analysed $\delta$\,Sct stars and second contains stars located only in the $\gamma$\,Dor area
($7100<$\,\teff\,$<7300$, KA16) to check whether there are any differences between their abundance patterns.
The comparisons are shown in Fig.\,\ref{figure12}.
It is clearly seen that $\delta$\,Sct stars in $\gamma$\,Dor area and the other $\delta$\,Sct stars have abundance patterns similar to $\gamma$\,Dor stars. 
However, abundance patterns of $\delta$\,Sct and $\gamma$\,Dor stars show small differences in comparison with the abundance patterns of non-pulsating stars. 
The \ion{Na}\ is overabundant in non-pulsating stars in comparison with $\delta$\,Sct and $\gamma$\,Dor variables. 
Additionally, abundances of \ion{Si}\ and \ion{Cu}\ are slightly lower in $\delta$\,Sct and $\gamma$\,Dor stars than in the non-pulsating stars.
On the other hand, \ion{Fe}\ abundances are similar in the investigated types of stars.
When we compare the abundance patterns of $\delta$\,Sct and $\gamma$\,Dor stars,
it turned out that \ion{Zn}\ abundance is lower in $\gamma$\,Dor stars than in $\delta$\,Sct stars,
and \ion{Sr}\ abundance is higher in $\delta$\,Sct stars than in the non-pulsating and $\gamma$\,Dor stars.
\citet{2008A&A...485..257F} found \ion{Y}\ and \ion{Ba}\ are over-abundant in $\delta$\,Sct stars in comparison to the non-pulsating stars.
However, we did not find any difference in abundances of these elements for $\delta$\,Sct and non-pulsating stars.

\section{conclusions}

This study presents the detailed spectroscopic investigation of a sample of $41$ $\delta$\,Sct stars. 
The initial atmospheric parameters (\teff\ and \logg) were derived from the photometric indices, SED and Balmer lines.
The accurate spectroscopic atmospheric parameters (\teff, \logg\ and $\xi$), \vsini\ values, and chemical abundances were obtained with the spectrum synthesis method. 

\teff\ and \logg\ values of $\delta$\,Sct stars were found in ranges of $6600-9400$\,K and $3.4-4.3$, respectively. 
The \vsini\ values were derived from $10$ to $222$\,km\,s$^{-1}$.
Additionally, these parameters of $\delta$\,Sct stars were compared with the parameters of $\gamma$\,Dor derived by KA16. 
As expected the average \teff\ value of $\delta$\,Sct stars is higher than the average \teff\ value of $\gamma$\,Dor stars.
We showed that $\delta$\,Sct stars are more evolved than the $\gamma$\,Dor stars. 
No significant difference was found between the average values and ranges of \vsini. 

The correlations between derived parameters and pulsation quantities were examined.
As shown by \citet{1990DSSN....2...13B}, the $P_{\rm puls}$ varies depending on the \logg, \teff, and bolometric magnitudes.
In our study, we found that a strong correlation between \teff\ and $P_{\rm puls}$ exists.
The \teff\,$-$\,$P_{\rm puls}$ correlation was also obtained for $\gamma$\,Dor stars (KA16) but this correlation was not as strong as for $\delta$\,Sct stars.
Additionally, although we did not find a significant \teff\,$-$\,$Amp$ correlation, it is obvious that the hotter stars in our sample have the lowest $Amp$ values. 
In the case of hotter stars, pulsation mechanism occurs very close to the stellar surface and it is not significantly effective to drive pulsations.
This could explain the lower $Amp$ values in hotter stars.

The \logg\,$-$\,$P_{\rm puls}$ relation for pulsating stars has been known before \citep[e.g.][]{1990ASPC...11..481C,1990DSSN....2...13B, 2017MNRAS.465.1181L}. 
However, because our sample consists of stars in a narrow \logg\ range, we did not find relations between $P_{\rm puls}$, $Amp$ and \logg.  
Similarly, no correlation between $\xi$ and $P_{\rm puls}$ was found for the analysed $\delta$\,Sct stars.
On the other hand, a weak positive correlation was obtained for $\gamma$\,Dor stars (KA16). 
Furthermore, the effect of \vsini\ on pulsation quantities was checked.
Although a strong negative correlation between \vsini\ and $P_{\rm puls}$ was obtained for $\gamma$\,Dor stars in the previous study (KA16),
the similar correlation was not obtained for $\delta$\,Sct stars analysed here. 
The $\delta$\,Sct stars pulsate in shorter periods and with higher $Amp$ in comparison with the $\gamma$\,Dor stars.
Because of the range of $P_{\rm puls}$ values of investigated $\delta$ Sct stars, the effect of \vsini\ on $P_{\rm puls}$ could not be obtained.
Additionally, weak negative correlations between \vsini\ and $Amp$ were found for both types of variables. 

We examined the positions of $\delta$\,Sct stars in $\log$\,\teff\,$-$\,\logg\ diagram and we conclude that most of our stars are located in $\delta$\,Sct instability strip.
Only three stars were found outside of their domain.
The stars located beyond the blue edge of $\delta$\,Sct domain, have the lowest $Amp$ values in comparison with the other analysed variables. 
Additionally, some of investigated stars are placed in an overlapping area of $\delta$\,Sct and $\gamma$\,Dor instability strips and the others are in $\delta$\,Sct domain.
When we compare pulsation properties of these two groups, we notice that the stars in overlapping area have higher $Amp$ values.
This result is in agreement with the \teff\,$-$\,$Amp$ relation shown in Fig.\,6.
The stars in $7100-7300$\,K area have higher $Amp$ values comparing the hotter ones.

Comparison of the abundance patterns of $\delta$\,Sct, $\gamma$\,Dor, and non-pulsating stars were considered. 
We found that $\delta$\,Sct stars have abundance pattern very similar to $\gamma$\,Dor stars.
However, \ion{Na}\ was obtained overabundant in non-pulsating stars in comparison with $\delta$\,Sct and $\gamma$\,Dor stars,
and \ion{Si}, \ion{Cu}\ were found underabundant in $\delta$\,Sct and $\gamma$\,Dor stars in comparison with the non-pulsating ones.
Additionally, \ion{Zn}\ is slightly less abundant in $\gamma$\,Dor stars comparing with $\delta$\,Sct stars,
and \ion{Sr}\ was obtained more abundant in $\delta$\,Sct stars than in the non-pulsating and $\gamma$\,Dor stars.
The \ion{Fe}\ abundance was determined to be almost the same in all types of stars. 
The suggested chemical differences between $\delta$\,Sct, $\gamma$\,Dor and non-pulsating stars can help us to 
understand why some stars in classical instability strip do not pulsate.
However, such a chemical abundance comparison needs a bigger sample.

Accurate atmospheric parameters and chemical abundance patterns of $\delta$ Sct variables were derived.
These parameters are important ingredients for a reliable seismic modelling of pulsating stars. 
Thus, the examination of the internal structure of stars in any evolutionary stage can be derived more accurately.
Additionally, obtained abundance differences between $\delta$\,Sct, $\gamma$\,Dor, and non-pulsating stars may give 
us a first approach of understanding why some stars located in the classical instability strip do not show pulsations.

\section*{Acknowledgments}
The authors would like to thank the reviewer for useful comments
and suggestions that helped to improve the publication.
This work has been partly supported by the Scientific and Technological Research Council of Turkey (TUBITAK) grant numbers 2214-A and 2211-C.
EN, MP, JM\.{Z} and KH acknowledges support from the NCN grant No. 2014/13/B/ST9/00902. 
The calculations have been carried out in Wroc{\l}aw Centre for Networking and Supercomputing (http://www.wcss.pl), grant No.\,214.
We are grateful to Dr. D. Shulyak for putting the code for calculating SEDs at our disposal. 
We thank to Dr. G. Catanzaro for putting the code for Balmer lines analysis at our disposal. 
This research has made use of the SIMBAD data base, operated at CDS, 
Strasbourq, France. This work has made use of data from 
the European Space Agency (ESA) mission Gaia (http://www.cosmos.esa.int/gaia), processed by the Gaia Data Processing 
and Analysis Consortium (DPAC, http://www.cosmos.esa.int/web/gaia/dpac/consortium). Funding for the DPAC has been 
provided by national institutions, in particular the institutions participating in the Gaia Multilateral Agreement.

%

\end{document}